\documentclass[fleqn,twoside]{article}
\usepackage{espcrc2}

\usepackage{graphicx}
\usepackage[figuresright]{rotating}

\newcommand{\AmS}{{\protect\the\textfont2
  A\kern-.1667em\lower.5ex\hbox{M}\kern-.125emS}}

\title{Small x evolution: BFKL vs Dipole Picture}

\author{G.P. Vacca \address[DPINFN]{Dipartimento di Fisica - Universit\`a
       di Bologna and INFN - sezione di Bologna \\
       Via Irnerio 46 , Bologna, Italy} 
           }
\begin{document}

\begin{abstract}
The relation between the two approaches is discussed both in the
linear and nonlinear regimes. It is connected to the gauge invariance and
M\"obius symmetry in LL perturbative QCD.
First corrections beyond the large $N_c$ approximation are discussed
for the nonlinear case.
\vspace{1pc}
\end{abstract}

\maketitle

\section{Introduction}
The study of high energy scattering processes where strong
interactions take place and where dynamics may be understood in terms
of high parton densities are under strong investigations since many years.
The most common framework of approximations for analytical
computations is based on the assumption of considering fixed $\alpha_s$.
In the diagrammatic technique of perturbative QCD one may study such
processes in the Leading Logarithmic approximation (multi Regge kinematics).
The first step is to consider the scattering amplitude of a projectile
on a target which, in the context of high energy factorization,
is constructed from the following ingredients:
(a) the external particle (taken colourless) impact factors which are
functions of the trasverse momenta of the gluon (reggeized) exchanged
in the $t$-channel, functions which vanish due to gauge invariance
when the momentum of any of the attached reggeized gluons goes to zero;
(b) the BFKL~\cite{BFKL} Green's function built from two interacting
$t$-channel gluons which defines an evolution in rapidity of the
scattered system.

Due to the requirements of unitarity in the $t$-channel one should
consider larger classes of $t$-channel states~\cite{Bartels,KPJ}
as well as verteces describing the transition between
them~\cite{Bartels} (``generalized'' BFKL approach). This fact is crucial and
unavoidable with the increasing of the rapidity intervals involved.
This fact appears consistent with the observation of a strong rise at small $x$
(or large rapidities) of the gluon distributions induced by the BFKL
Green's function and the cross sections, with a power behavior in the
energy which contraddicts the Froissart bound.
Moreover there is also a connection with the attempts to find an effective 2+1
dimensional field theory for the QCD in the Regge limit which may be
useful to reveal links between QCD, SUSY QCD and finally string theories
from the ADS/CFT correspondence.

The main point considered in the following is to recalling
how is the relation between the ``generalized'' BFKL approach,
based on resummation of Feynman diagrams, and the dipole picture
approach~\cite{dipoles}, whithin which a first description of
effective nonlinear phenomena in QCD at small $x$ (in
the large $N_c$ limit approximation) was given~\cite{BLV2}.
This relation is understood in terms of gauge freedom and the large
$N_c$ limit approximation. A first contribution beyond the large $N_c$
approximation is discussed.    
\section{Linear Evolution: BFKL and Colour Dipole Physics}
The main result of the BFKL approach is that the leading
contribution to the cross section can be written as an integral in the
transverse space
\begin{eqnarray}
\sigma \simeq \int d \mu_T \, \Phi_1 \, G(y) \, \Phi_2 \!\!\!\!\!\!\!\!&&, 
\frac{\partial}{\partial y}\,G=\delta +\frac{\bar{\alpha}_s}{2}K\, G
\, , \nonumber \\
&& K=\omega_1 +\omega_2 +V \, ,
\label{HEF}
\end{eqnarray}
where $y$ is the rapidity which plays the role of an evolution
parameter and $d\mu_T$ is the measure in the transverse space and $K$
is the BFKL kernel. The perturbative kernel $K$ is getting contributions from
virtual ($\omega_i$) and real (V is related to the square
of the Lipatov vertex for gluon production in the LL approximation)
and is free of I.R. divergences in the colour siglet case.
The Green's function $G$ defined above may be written in terms
of the spectral basis of the operator $K$, solution of the so called
BFKL equation.
The interacting two reggeized gluon system in a colour singlet state is
known as the perturbative BFKL Pomeron and the eigenvalues of the
kernel are related to its intercept. 
Let us note that the reggeized gluons are defined in a selfconsistent
way looking at the BFKL kernel properties in the colour octet state in the
$t$-channel (bootstrap property).

The above mentioned property of the impact factors of colourless
external particles (related to the gauge invariance) gives
the freedom to look for different possible representations of the
space of functions of the impact factors and for the domain of the
operator $K$.
In particular one may consider the M\"obius representation when
a space of functions $f(\rho_1,\rho_2)$ such that
$f(\rho,\rho)=0$ is considered.
In such a case the BFKL equation is invariant under the M\"obius,
conformal, transformations and the BFKL kernel
posses also the holomorphic separability property,
which means that in the coordinate space, when defining a complex
$\rho=\rho_x + i \rho_y$, the kernel can be decomposed in a sum
$K=h+h^*$, where $h=h(\rho_i)$.

Starting from the Feynman diagrams derivation, in momentum space,
the BFKL kernel is directly seen as a pseudodifferential operators
acting on functions with complex variables:
\begin{eqnarray}
\!\!\!\!\!\!\!\!\!\!&-&\!\!\!\!K=H_{12}=
\ln \,\left| p_{1}\right| ^{2}+\ln \,\left| p_{2}\right| ^{2}-4\psi(1)+
\nonumber \\
&&\!\!\!\!\!\frac{1
}{p_{1}p_{2}^{\ast }}\ln \left| \rho _{12}\right| ^{2}\,p_{1}p_{2}^{\ast
}+\frac{1}{p_{1}^{\ast }p_{2}}\ln \left| \rho _{12}\right|
^{2}\,p_{1}^{\ast }p_{2}\,.
\end{eqnarray}
In the M\"obius representation different transformations may be
derived, among which we find the one tipical of the colour dipole
picture, with a purely integral operator form:
\begin{equation}
\!\!K N\!=\!\!
\int\! \frac{d^{2}\rho _{3}}{\pi }\,\frac{\left| \rho _{12}\right| ^{2}}{
\left| \rho _{13}\right| ^{2}\left| \rho _{23}\right| ^{2}}\,(
N_{\rho_{1},\rho_{3}}+N_{
\rho_{3},\rho_{2}}
-N_{\rho_{1},\rho_{2}}) \,.
\end{equation}
In this formalism the cross section is written as
\begin{equation}
\sigma \simeq
 \int d^{2}\rho_{1}d^{2}\rho_{2}\,\int_{0}^{1}dx\,\left|
\psi(\rho_1, \rho_2;x)\right| ^{2}\,N(\rho_{1},\rho_{2}; y)\,
\end{equation}
and the relation with the form in Eq. (\ref{HEF}) can be studied.
For a virtual photon this translates in finding the relation between
the impact factor $\Phi_1$ and its wave function $\psi$~\cite{BLV2}.
One has
$\Phi_1=\int dx |\psi|^2 \theta_{IR}$, where $\theta_{IR}$ are the
phase factors which describes the four ways the two gluons attach to
the $q\bar{q}$ pair ($\theta_{IR}\to 0$ is related to gauge
invariance) and give the freedom to add any term with no overlapping
support. 
Using such a gauge freedom one may write, in terms 
of an operator which projects onto the
M\"obius space of function, $\theta^{UV}$, such that
$\theta^{UV} f_{\rho_1,\rho_2}=f_{\rho_1,\rho_2}-1/2 f_{\rho_1,\rho_1}
-1/2 f_{\rho_2,\rho_2}$:
\begin{eqnarray}
\sigma \simeq \!\!\!\!\!\!\!\!\!\! &&\Phi_1 \otimes G \otimes \phi_2=
\int dx |\psi|^2 \theta_{IR} \otimes G \otimes \Phi_2 =\nonumber \\
&&\int dx |\psi|^2 \otimes \theta^{UV} G \otimes \Phi_2 = 
\int dx |\psi|^2 \otimes N \,.
\end{eqnarray}
As discussed in the introduction one needs to study in the LL
approximation the linear evolution of more
complicated systems, with many reggeized gluons (more than 2) in the
$t$-channel.
The homogeneous equation (BKP)~\cite{Bartels,KPJ}
for n-gluon colour singlet states are governed by a kernel $K_n$
which is a sum of 2-gluon kernels defining a Green's function:
\begin{equation}
\frac{\partial}{\partial y}\,G_n=\delta +\frac{\bar{\alpha}_s}{2} K_n\, G_n\,.
\end{equation}
Let us note that the n-gluon impact factors must have the property of vanishing
for a zero gluon momentum, but, nevertheless, the gauge freedom
does not allow to restrict all the solution to a space of function
which are null when two coordinates coincide.
The simplest case appears at the $3$ gluon states, since the leading
intercept Odderon solution~\cite{odderon} does not have
this property.
Anyway there exist interesting states which belong to a (generalized)
M\"obius representation which are dynamically compatible with the BFKL
evolution and will be considered in the following step towards
the unitarization.

\section{Non Linear Evolution: Leading Large $N_c$ and a Step Beyond}
On considering systems with different number
of gluons in the $t$-channel it is possible to organize a hierarchy of
an infinite number of coupled linear equations~\cite{Bartels}.
In particular these set of equations has been written and analyzed
explicitely in a systematic way for the systems of up to $6$
gluons~\cite{hierarchy4,hierarchy6}.  
In the $4$ gluon system one may extract an effective vertex
$V_{2\to4}$ which
defines the transition between $2$ and $4$ reggeized gluons. Any
solution can be decomposed~\cite{hierarchy4} in the sum of two terms,
$D_4^R$ and $D_4^I$, the latter being related to such a transition,
followed by an evolution governed by $G_4$.
The former term is governed by the gluon reggeization and different
choices can be made in the large $N_c$ limit~\cite{BV,BBV}.
In the large $N_c$ limit, when only colour planar diagrams dominates,
the effective vertex $V_{2\to4}$ becomes a simpler vertex which
decribes the splitting of a BFKL Pomeron into two. 
Also in the $6$ gluon system one may observe a particular contribution
to the solution which is an iteration of two splitting of
the kind $V_{2\to4}$ in sequence in rapidity, each followed by the
BFKL evolution.

The diagrams with successive splitting in rapidity, denoted fan
diagrams, have been defined in the past in a different context~\cite{GLR}. 
Therefore one is tempted to consider all the diagrams, where
splitting in sequence are present, and to define an object which
describes a full resummation.
This approach has led to derive in a special case~\cite{num} the BK equation,
previously obtained by the resummation of colour dipole splitting in nuclear
targets~\cite{BK} in the limit $N_c\to \infty$. A more general
investigation also beyond such an approximation can be found in~\cite{BLV2}.

As usual, the non linear evolution appears when one is insisting in
defining an approximation of the full system, which is governed by linear
equations, in terms of a single smaller object and neglecting all kind
of higher correlations.

\begin{figure}[hbtp]
  \begin{center}
    \includegraphics[width=0.8 \linewidth]{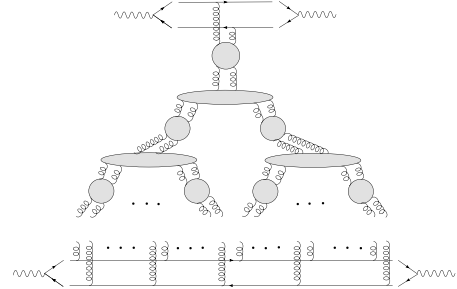}
    \caption{Fan diagrams which are resummed, with the coupling of the
    gluons to the quark lines understood in all possible ways.}
    \label{Fig:ex3}
  \end{center}
\end{figure}

As a first step the leading fan structure is extracted on considering
the substitution $G_4 \to G_2 \otimes G_2$ so that one may write
\begin{equation}
\frac{\partial}{\partial y} \Psi=
\frac{\bar{\alpha}_s}{2} K \, \Psi - \bar{\alpha}_s^2 {\cal V}
\otimes \Psi \Psi \, ,
\label{eqfan}
\end{equation}
which resums the fan diagrams of Fig.~\ref{Fig:ex3}.
With ${\cal V}$ we have denoted the effective transition vertex.
The choice of using the M\"obius representation greatly simplifies the
calculations and Eq. (\ref{eqfan}), after defining
$N_{\rho_1,\rho_2}=8\pi \alpha_s \theta^{UV} \Psi_{\rho_1,\rho_2}$,
reads exactly like the BK~\cite{BK} equation.

This equation has been derived in the large $N_c$ limit approximation.
In order to write an extension in the next-to-leading
$1/N_c$ approximation, there are two sources of corrections:
the nonplanar term, subleading in $1/N_c$, in the effective
vertex and the $4$-gluon Green's function. In particular for the latter
it is convenient to study $N_{4\, \rho_1,\rho_2;\rho_3,\rho_4}=
N_{\rho_1,\rho_2}N_{\rho_3,\rho_4}+\Delta
N_{4\, \rho_1,\rho_2;\rho_3,\rho_4}$, which evolves according to
$\frac{\partial}{\partial y}\,N_4=\frac{\bar{\alpha}_s}{2} K_4\, N_4$.
Therefore it is possible to write a system of two coupled
equations~\cite{BLV2}:
\begin{eqnarray}
\!\!\!\!\!\!\!\!\!\!\!\!\!\!\!\!\!\!\!\!\!\!
\frac{d}{dy}N_{x,y}\!\!\!\!\!\!\!\!\!\!\!&&=
\bar{\alpha}_s
\!\!\int \!\!\frac{d^{2}z}{2\pi }
C_{x,y,z}
\Bigl[ N_{x,z}\!+\!N_{z,y}\!-\!N_{x,y}
\!-\!N_{x,z}N_{z,y}\nonumber \\
&&-{\Delta N_{4\, x,z;y,z}
- \frac{1}{2}\frac{1}{N_c^2-1}
\left( N_{x,z}+N_{z,y}-N_{x,y}\right)^2} \Bigr]\, ,
\nonumber \\
&&\!\!\!\!\!\!\!\!\!\!
\frac{d}{dy} \Delta N_{4\, \rho_1,\rho_2;\rho_3,\rho_4}
= \, \frac{\bar{\alpha}_s}{2}\Bigl[ \left(K_{12} + K_{34} \right)
\Delta N_{4\, \rho_1,\rho_2;\rho_3,\rho_4} + \nonumber \\
&&\!\!\!\!\!\!\!\!\!\!\!\!\!\!\!
\frac{1}{N_c^2 -1} \left(K_{12} + K_{34} \right)
\left(N_{\rho_1,\rho_3} N_{\rho_2,\rho_4} +
N_{\rho_1,\rho_4} N_{\rho_2, \rho_3} \right)\Bigr] \,,
\end{eqnarray}
where $C_{x,y,z}=\frac{| x-y| ^{2}}{|x-z| ^{2}| y-z| ^{2}}$ and on
considering only the first line of the first equation one reobtains
the BK case.
\section{Conclusions}
A summary of a recent investigation\cite{BLV2} of the link between
the Feynman diagram analysis a la BFKL and the dipole picture has been given.
This link has also permitted to write an explicit correction of the non
linear evolution equation for colour dipoles beyond the large $N_c$ limit
approximation. Quantitative but not qualitative changes are expected
in the evolution. 
In such a framework are still missing the NLL corrections (in
$\alpha_s$) and any kind of ``loop'' corrections in the $t$-channel,
clearly required by any effective description of high energy QCD.
Much more theoretical work is deserved for the future.


\begin{thebibliography}{9}
 \bibitem{BFKL}
L.~N.~Lipatov, Sov.\ J.\ Nucl.\ Phys.\ {\bf 23} (1976) 338; 
V.~S.~Fadin, E.~A.~Kuraev and L.~N.~Lipatov, Phys.\ Lett.\ B 
{\bf 60} (1975) 50;
I.~I.~Balitsky and L.~N.~Lipatov, Sov.\ J.\ Nucl.\ Phys.\ {\bf 28} (1978) 822;
\ JETP Lett.\ {\bf 30} (1979) 355.
\bibitem{Bartels}
J.~Bartels, Nucl.\ Phys.\ B {\bf 175} (1980) 365;
 \bibitem{KPJ}
J.~Kwiecinski and M.~Praszalowicz, Phys.\ Lett.\ B {\bf 94} (1980) 413;
T.~Jaroszewicz, Acta Phys.\ Polon.\ B {\bf 11} (1980) 965.

\bibitem{dipoles}
A.H.Mueller, Nucl. Phys. {\bf B 415} (1994) 373; {\bf B 437}
(1995) 107;
A.~H.~Mueller and B.~Patel, Nucl.\ Phys.\ B {\bf 425} (1994) 471.

\bibitem{BLV2}
J.~Bartels, L.~N.~Lipatov and G.~P.~Vacca, arXiv:hep-ph/0404110.

\bibitem{odderon}
J.~Bartels, L.~N.~Lipatov and G.~P.~Vacca,
Phys.\ Lett.\ B {\bf 477} (2000) 178
[arXiv:hep-ph/9912423].

 \bibitem{hierarchy4}
J.~Bartels and M.~Wusthoff, Z.\ Phys.\ C {\bf 66} (1995) 157.

 \bibitem{hierarchy6}
J. Bartels and C. Ewerz, JHEP {\bf 9909} (1999) 026.

 \bibitem{BV}
M.~A.~Braun and G.~P.~Vacca, Eur.\ Phys.\ J.\ C {\bf 6} (1999) 147.
G.~P.~Vacca, PhD thesis, arXiv:hep-ph/9803283.

\bibitem{BBV}
J.~Bartels, M.~Braun and G.~P.~Vacca,
arXiv:hep-ph/0412218.

\bibitem{GLR}
L.~V.~Gribov, E.~M.~Levin and M.~G.~Ryskin,
Phys.\ Rept.\  {\bf 100} (1983) 1.

\bibitem{num}
M.~Braun, Eur.\ Phys.\ J.\ C {\bf 16} (2000) 337.

\bibitem{BK}
I.~I.~Balitsky, Nucl.\ Phys.\ B {\bf 463} (1996) 99; \\
Y.~V.~Kovchegov, Phys.\ Rev.\ D {\bf 60} (1999) 034008;
ibid. Phys.\ Rev.\ D {\bf 61} (2000) 074018. 

\end{thebibliography}
\end{document}